\documentclass[letter,fleqn]{cas-sc}
\usepackage{setspace}
\usepackage[authoryear]{natbib}
\usepackage{amssymb, amsmath}

\newcommand{\dpm}[2]{\substack{+#2 \\ -#1}}

\begin{document}
\let\WriteBookmarks\relax
\def\floatpagepagefraction{1}
\def\textpagefraction{.001}
\shorttitle{Improved Automated Crater Detection}
\shortauthors{Lee and Hogan}

\title [mode = title]{Automated crater detection with human level performance}

\author[1]{Christopher Lee}[orcid=0000-0003-0029-5278, role=Researcher]

\cormark[1]
\fnmark[1]
\ead{clee@atmosp.physics.utoronto.ca}
\ead[url]{www.christopherlee.co.uk, clee@atmosp.physics.utoronto.ca}

\credit{Conceptualization of the study, methodology, original software, writing}
\address[1]{Department of Physics\\
University of Toronto\\
60 St. George St,
Toronto,
M5S 1A7,
Canada
}

\author[1]{James Hogan}[orcid=0000-0003-4694-5176, role=Researcher]
\fnmark[1]
\ead{james.hogan@mail.utoronto.ca}

\credit{Algorithm development, optimization, methods section}

\cortext[cor1]{Corresponding Author}

\nonumnote{\printcredits}
\begin{abstract}
Crater cataloging is an important yet time--consuming part of geological mapping. We present an automated Crater Detection Algorithm (CDA) that is competitive with expert--human researchers and hundreds of times faster. 

The CDA uses multiple neural networks to process digital terrain model and thermal infra--red imagery to identify and locate craters across the surface of Mars. We use additional post-processing filters to refine and remove potential false crater detections, improving our precision and recall by 10\% compared to \citet{Lee2019}. We now find 80\% of known craters above 3km in diameter, and identify 7,000 potentially new craters (13\% of the identified craters). The median differences between our catalog and other independent catalogs is 2--4\% in location and diameter, in--line with other inter--catalog comparisons. 

The CDA has been used to process global terrain maps and infra--red imagery for Mars, and the software and generated global catalog are available at \url{https://doi.org/10.5683/SP2/CFUNII}.

\end{abstract}



\maketitle

\doublespacing
\label{sec:introduction}
\section{Introduction}
\label{sub:motivation}

Crater populations and distributions provide critical information on the history and evolution of Solar system bodies. Regional variation in crater distribution can be used to constrain stratigraphy and geologic processes \citep[e.g.,][]{Cintala1976,Barlow2003}, and populations are used in estimated the age of surface features \citep[e.g.,][]{Arvidson1974, Soderblom1974, Kinczyk2020, Palucis2020}. 

Measuring craters has historically required manually measuring the size and location of craters in images of the surface. Early work on crater mapping required manual annotation of printed maps\citep{Barlow1988}. Advances in imagery products and computational tools allowed more detailed work using digital mapping \citep{Robbins2012} and computer aided detection of craters using Crater Detection Algorithms (CDAs, e.g., \citet{Stepinski2009}, \citet{Di2014}, \citet{Pedrosa2017}). These early CDAs uses manually designed rules to identify craters, but recent advances in Deep Learning\citep{Goodfellow2016} permit algorithms that learn rules using the data itself and a \textit{ground--truth} catalog of known craters. These neural network based CDAs have shown performance approaching expert human level \citep{Silburt2019,Lee2019}, but have fallen short of matching or exceeding expert performance when compared against independent expert generated catalogs \citep[e.g.,][]{Salamuniccar2012}.


In this work, we develop a CDA that is capable of finding near--circular craters from orbital imagery and terrain data with expert human level performance while being hundreds of times faster. We improve on the \citet{Lee2019} CDA using additional datasets and post--processing to remove false positives and less confident crater detections to find almost 60,000 craters from 3km to 450km in diameter. The CDA workflow described in the next section can be used on imagery data of the same type with little or no training, and can be trained to work on new data from visible and ultra--violet wavelengths. New post--processing stages merge and de--duplicate global catalogs generated from independent sources to improve the accuracy and completeness of the final catalogs.

In this paper we describe the structure and workflow of the algorithm in section \ref{sec:methods}, including the crater detection neural networks and the post-processing filters that improve the precision of final catalogs. In section \ref{sec:results_and_discussion} we use this CDA to process DTM and Infra--red imagery data from Mars to generate a new catalog of craters above 3km in diameter, and we compare this new catalog to two existing crater catalogs for Mars \citep{Robbins2012a,Salamuniccar2012} to show that the CDA catalog is statistically competitive with the human generated catalogs. Finally, in section \ref{sec:conclusions}, we provide our conclusions and possible future steps to improve the CDA and generated catalogs.
 

\subsection{Previous Work}\label{previous-work}

The \citet{Lee2019} CDA (hereafter L19) used the DeepMoon CDA developed by \citet{Silburt2019} to find craters in a digital terrain model (DTM) of Mars. This CDA worked by finding the rims of craters based on altitude, and learned by comparing images to the location of known craters from a \textit{ground--truth} catalog\citep{Robbins2012}. The CDA found almost 55,000 craters, larger than 4km in diameter, distributed across the planet with a precision and accuracy of 75\%. Approximately 13,000 `new' craters were proposed by the CDA and a similar number of known craters were missing in the CDA catalog. A large fraction of these `new' crater candidates are anomalous detections of small craters in the DTM data, with some larger genuine features from canyon rims and volcanic paterae. Under similar conditions, the \citet{Salamuniccar2012} catalog lists almost 72\% of the same craters from \citet{Robbins2012} but suggests less than 5,000 `new' craters.

Other automatically--generated crater catalogs exist for Mars. \citet{Stepinski2009b} developed a catalog using an automated CDA (following \citet{Bue2006}) in combination with statistical methods to combine disparate datasets into a larger catalog. \citet{Di2014} developed an automated CDA based on correlation methods, but only tested the algorithm on a small region of the planet. In both examples, the reported precision fell much lower than either L19 or \citet{Salamuniccar2012} in comparison to \citet{Robbins2012}, but the limited area tested makes comparison difficult (\citet{Lee2019} includes a quantitative comparison of these limited area tests). Similar work has been performed for Lunar craters \citep[e.g.,][]{Zuo2016} and for the Earth \citep[e.g.,][]{Krogli2010}.

Other groups have also created neural network based CDAs. \citet{Silburt2019} developed the UNET \citep{Ronneberger2015} based \textit{DeepMoon} CDA to detect craters on the Moon, which was subsequently modified in \citet{Lee2019}. \citet{Wronkiewicz2018} used the YOLO \citep{Redmon2016} neural network for Mars craters, and the results from this CDA in Jezero crater are included in the JMars package (\url{https://jmars.asu.edu/}). \citet{Lee2018a} used a MaskRCNN\citep{He2017} algorithm to find circular and elliptical craters for Mars and Pluto, \citet{Ali-Dib2019} used a similar network to find elliptical craters on the Moon.

Each of these CDAs identified and located craters much faster than human cataloging efforts. For example, the L19 CDA processes a global Mars DTM down to 4km crater diameters in 24 hours on a modest workstation, using 150,000 images to identify 55,000 craters. However, since the CDA catalog contains a large number of new and missed crater detection additional work is required to filter out spurious craters and complete the catalog.

The CDA developed in this work improves upon L19 by reducing by one--fifth the number of missed craters, and by halving the number of mistakenly identified craters \citep[compared to][]{Salamuniccar2012}. 


\section{Methods and Data}
\label{sec:methods}
\label{sub:workflow}

\begin{figure*}[ht]
    \centering
    \includegraphics[width=\textwidth]{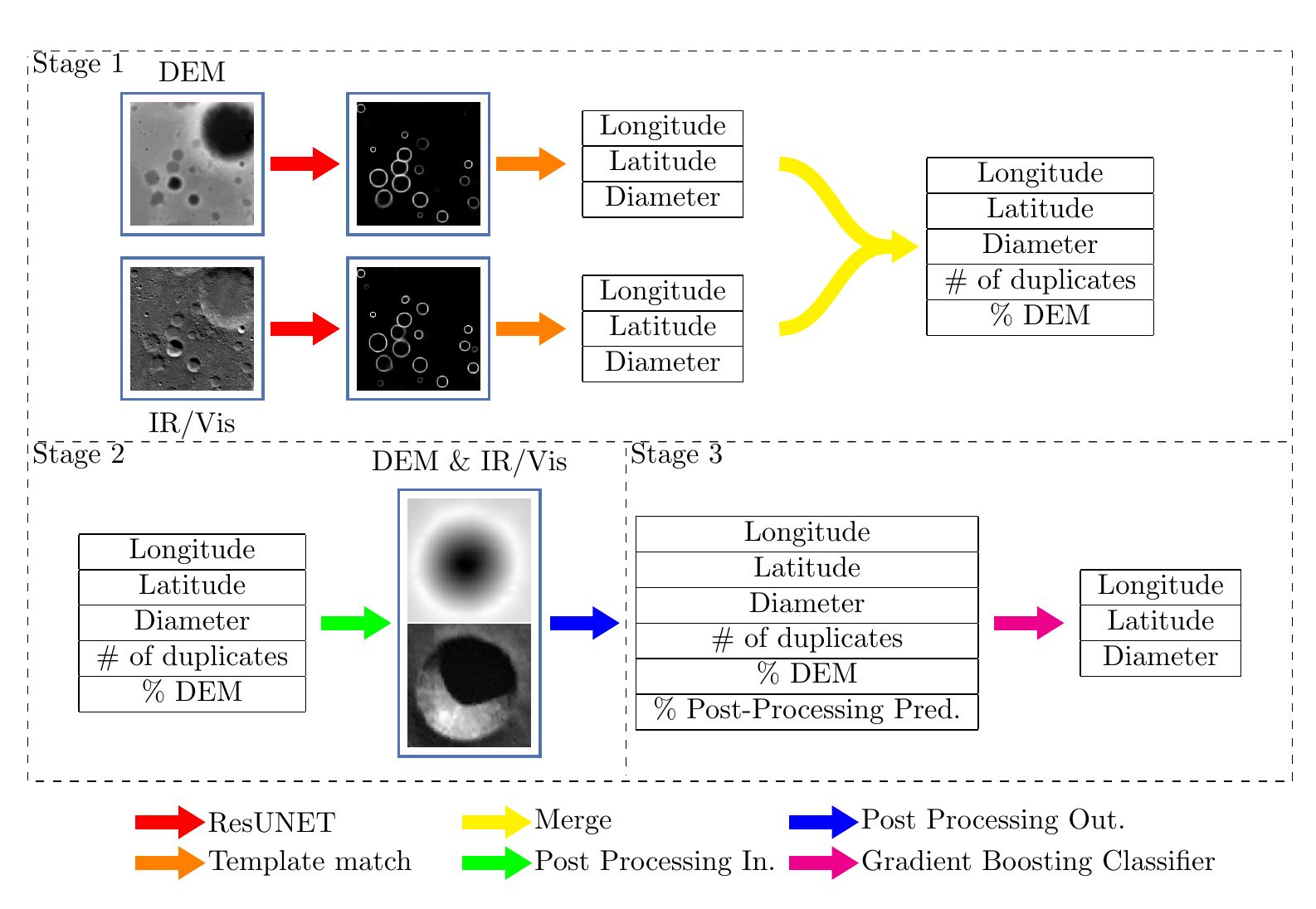}
    \caption{\label{figure:workflow} Workflow of the CDA. In stage 1 the ResUNETs convert DTM or IR images into synthetic images of the crater rims, followed by the template matching to determine the size and location of the craters. The crater datasets are then merged, removing any duplicates and counting the fraction of craters from each source. In stage 2 the global crater dataset is used to generate images of single craters, centred in the images and the post--processing network is used to provide confidence values. In stage 3 the Gradient Boosting filter is used to remove the weaker crater predictions.}
\end{figure*}

Our Crater Detection Algorithm (CDA) automates the process of finding and characterizing crater-like features on the surface of a rocky planet. This involves using multiple processing stages to identify crater candidates, confirm crater identifications, and finally output a crater catalog based on the information collected in the CDA. The workflow of the CDA, as used in this work, is shown in Figure \ref{figure:workflow}.

In stage 1 of the CDA, we use a ResUNET\citep{Zhang2018a} neural network to find circular crater features in global imagery or terrain model datasets (section \ref{sub:resunet}). For memory and performance considerations, we split the global dataset into smaller images sampled across the planet at various resolutions, and process each image using the ResUNET neural network to identify circular features.

In the workflow shown in Figure \ref{figure:workflow}, we use two datasets in the stage 1 of the CDA (section \ref{sub:resunet}). A global daytime Infra-Red (IR) map produced from observations by the Thermal Emission Imaging System instrument (THEMIS) onboard Mars Odyssey \citep{Edwards2011}, and the combined Digital Terrain Model (DTM) dataset generated from the Mars Orbiter Laser Altimeter (MOLA) and the High Resolution Stereo Camera (HRSC) \citep{Fergason2018}. Each global map is processed independently using a ResUNET trained to process that data, resulting in two crater catalogs. We then merge the catalogs generated by the ResUNETs into a single catalog, and duplicate craters are counted and merged. In the process of merging, we calculate the average crater location and size (and associated uncertainties) from the various detections of each crater and record the number of times each ResUNET found a crater.

In stage 2 of the CDA(section \ref{sub:classifier}), we use a \textit{classifier} neural network to calculate a `crater confidence' value for each crater in the catalog. This network uses \textit{both} the optical and terrain data to determine whether a crater is well-resolved in the dataset. The probabilities generated by this CDA are stored for later use but not used for filtering at this stage .

In stage 3 of the CDA(section \ref{sub:catalog-refinement}), we use the results from stage 1 and stage 2 to filter out false crater identifications and further refine the catalog, using a Gradient Boosting algorithm \citep[XGBoost,][]{Pedregosa} to filter out the false--positive identifications by considering diameter and duplicate count (from stage 1), and classification probability (from stage 2). At the end of this stage, the crater catalog contains the location, diameter, and associated uncertainties for each detected crater. The  catalog contains sufficient information to plot the craters using the JMars mapping tool (\url{https://jmars.asu.edu/}) that already includes various human and computer generated crater catalogs for Mars.

\subsection{Data}
The neural networks described in this section are flexible and can be trained to use different datasets that contain appropriate images of craters. In this work we use the DTM dataset from HRSC/MOLA\citep{Fergason2018} and the daytime infra--red (IR) imagery data from THEMIS\citep{Edwards2011}. The DTM product is provided as a single file covering the globe on an equirectangular grid with a stated scale of 200m/pixel. The daytime IR data is provided as a single equirectangular grid with a stated resolution of 100m/pixel. Both datasets are created by the satellite mission teams and registered to the Mars datum. We make no attempt to correct misalignments in these datasets. While we have used DTM and IR data in this work, it is possible to replace either dataset by (e.g.) visible imagery or add additional datasets to the workflow, providing craters can be identified in the images.

When finding the location and size of the craters the neural networks and template matching code reports locations in pixels using floating point numbers, providing a numerical uncertainty of 0.5 pixels in each reported number. This uncertainty is smaller than the uncertainty in defining a crater 'rim' in the images, and is comparable to the median difference between the CDA and the ground--truth dataset (5\% uncertainty at most, compared to a median difference of 4\% from Table \ref{table:results}).

In typical images, we find craters between 10 and 40 pixels in diameters at all locations. For the DTM data, this implies a lower diameter of $4\pm 0.2$km with location uncertainty of $0.2$km, and an upper diameter of around $500\pm6$km with a location uncertainty of $6$km. For the IR data, the lower limits are half the equivalent values from the DTM because of the higher resolution of the IR data. The upper detectable crater size in IR data is lower than 500km because the larger craters are not clean features in the infra--red data.

For the results discussed in section \ref{sec:results_and_discussion}, we found craters using the workflow by generating 272,370 image swatches at sizes from 1.5 degrees to 30 degrees in longitudinal extent covering the surface of Mars. We also generated an additional 1,292,726 images from the IR data only with image sizes from 0.5 to 1.3 degrees to exploit the higher resolution of the IR data. Results from these images are labelled as IRH in the following text. 

\subsection{Stage 1: ResUNET} 
\label{sub:resunet}

\begin{figure*}[ht]
    \includegraphics[width=\textwidth]{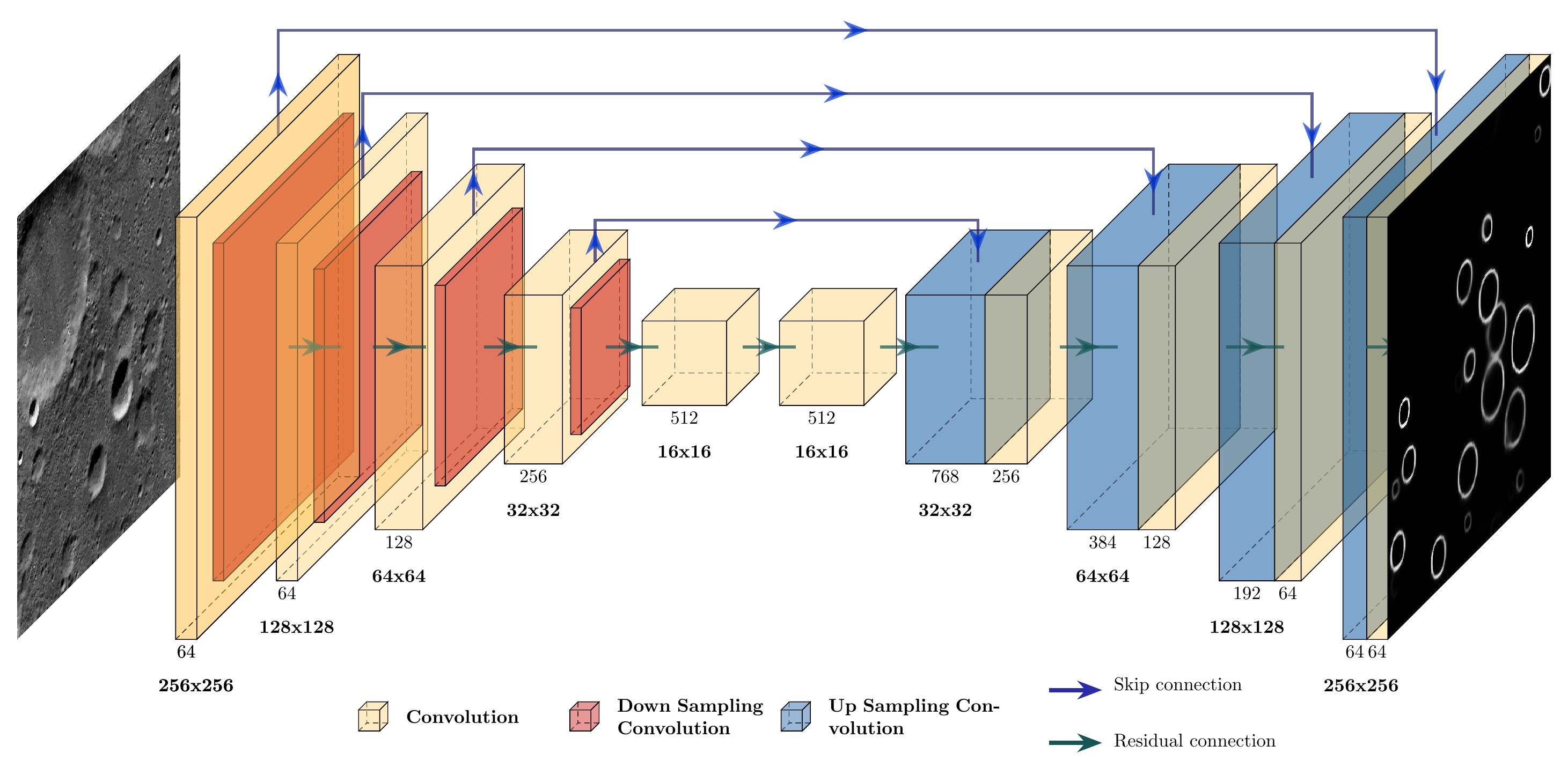}
    \caption{Our adaptation of the Residual UNET architecture using Keras \citep{Chollet2015}.  The layer types are shown in the figure legend. All convolution layers use a 2D convolution with a 3x3 kernel, and include batch normalization and use a rectilinear activation function. Residual connections carry the input to output changefrom each layer to the next layer, as in \citep{Fergason2018}.}
    \label{figure:resunet}
\end{figure*}

The first stage of the CDA uses a residual neural network, or ResUNET \citep{Zhang2018a}, to identify circular features in the input images that are highly correlated with desired features in the output images (i.e., crater rims). The ResUNET architecture is similar to the L19 network architecture \citep{Ronneberger2015}, but features additional internal connections that reduce training time and improve performance. Our ResUNET configuration is shown in Figure \ref{figure:resunet}.

The ResUNET uses an \emph{encoder} branch (the descending branch on the left) and a \emph{decoder} branch (the ascending branch on the right). The \emph{encoder} branch takes the original image and compresses the information down to a series of feature locations (i.e., crater locations) at the `bottom' of the network. The \emph{decoder} branch then constructs a new image by combining the feature locations with the original image data constructing an image with highlighted crater rims, while removing noise from the images.

We train the ResUNET following the same methodology as in L19. We generate forty thousand images with random locations across the planet and with a log--normal distribution of image sizes. The images are orthographically projected and presented to the ResUNET along with a \emph{target} image generated from a ground--truth crater catalog \citep{Robbins2012}. We train the network over many iterations until the performance of the network, measured using the Dice loss algorithm \citep{Milletari2016}, stops improving. We retain an additional 10,000 randomly generated images in a validation dataset to adjust external \textit{hyper parameters}.

As in L19, the ResUNET outputs synthetic images with the crater rims marked (shown in Figure \ref{figure:workflow}). We then convert each greyscale image to black and white, and use a circular template matching algorithm\citep{VanderWalt2014} to identify all circles with a radius of between 5 and 40 pixels that are a fraction $t_\mathrm{template}$ of a complete circle. This processes finds all circles in the image, with some duplicates at consecutive crater sizes and locations (e.g., within a pixel or two in diameter or location). We remove these duplicate craters in the image by merging craters that are within a radius threshold ($t_\mathrm{rad}$) and location threshold ($t_\mathrm{distance}$) of another crater, and choosing the crater properties that maximize the correlation with the target circle. That is, duplicates satisfy the following equations

\begin{eqnarray}
\nonumber \frac{|r_1-r_2|}{\min(r_1,r_2)} &\le& t_\mathrm{rad},\\
\sqrt{\frac{(x_1-x_2)^2 + (y_1-y_2)^2}{\min(r_1,r_2)^2}}  &\le& t_\mathrm{distance}
\end{eqnarray}

where the position and radius of the candidate craters are given by $(x_i,y_i)$ and $r_i$. We determined the threshold values for this work by maximizing the recall of the ResUNET on the 10,000 image validation dataset, and use the values $t_\mathrm{template}=0.475$, $t_\mathrm{target}=0.4$, $t_\mathrm{rad}=0.3$, $t_\mathrm{distance}=0.3$.

The output from the ResUNET is a crater candidate list with no per--image duplicates but possible global duplicates. We do not remove the global duplicates at this stage, instead using them to refine the location of each crater and estimate uncertainties in the following stages.

After the ResUNET processing is complete for two or more input datasets, we merge the individual catalogs into a global catalog. We merge multiple identifications of the same crater (craters within 25\% in size and relative position of each other) and count the number of times the crater is found in each dataset. We record the duplicate count, average location and diameter of each crater, and the standard deviation in location and size for each crater in the stage 1 catalog.

An alternative stage 1 network was tested where the two image datasets are used simultaneously in a two channel ResUNET. This network did not improve on the L19 as the DTM and IR 1--channel ResUNETs perform well at different spatial scales, and thus are better combined after the initial crater detection. The final improvements we found (i.e., stage 2 and 3) were gained by removing false positives using the independent datasets. We also limited the input data to image scales where both datasets could contribute to avoid biasing the results towards one source of data. The individual ResUNET stages here can be trained to work on high resolution data for crater identification and other feature mapping \citep[e.g.,][]{Palafox2017,Wronkiewicz2018, Lee2019}.

\subsection{Stage 2: Classifier} 
\label{sub:classifier}

In stage 2, we use a \textit{classifier network} to derive a `crater confidence' estimate for all of the candidates in the stage 1 catalog. The classifier network has a similar structure to the \textit{encoder} branch of the ResUNET in stage 1, but converts a 2 channel (IR and DTM) image into a single number representing the crater confidence value.

To process each crater candidate, the classifier network uses the location and diameter of each candidate to generate IR and DTM images centered on the crater candidate with a 20\% border around the crater. The network then assigns a confidence value to the crater candidate based using both the IR and DTM images together.

 To train the classifier, we take 1,000 random crater samples and 1,000 random non--crater samples (that might contain a crater) and generate images containing each target crater centered in the image. We then train the network to assign high confidence values to the crater images and low confidence values to the non--crater images, without specifically training it to identify circles, edges, rims, or shadows. Once trained, we use the classifier network to process the entire stage 1 catalog and add the crater confidence value for each crater to the catalog. We do not remove craters at this stage.


\subsection{Stage 3: Catalog Refinement}\label{sub:catalog-refinement}
In the final stage of the CDA we filter craters from the catalog using the data collected in the previous stages. We train the filter by sampling 1,000 crater candidates from the CDA catalog and identify which craters match the \textit{ground--truth} catalog. We then use a \textit{Gradient Boosting Classifier} model \citep[XGBoost, ][]{Pedregosa} to predict and remove false positive craters based on the initial 1,000 samples.

The XGBoost algorithm learns to map the detected features (diameter, DTM duplicate count, total duplicate count, likelihood) into a binary crater/non--crater value, and we remove all `non--craters' from the final catalog.  In typical usage, the XGBoost model favors high duplicate count, equal representation in IR and DTM images, and a high classifier confidence. These features are automatically learned but likely represent the idea that a real crater should be identifiable at many image scales in both IR and DTM images, and should look like other craters. We include the diameter of the crater in this model because very large and very small craters tend to have low duplicate counts and are not found in both IR and DTM data equally well. At the end of this stage, the crater catalog has been globally deduplicated, filtered for false positives, and contains crater location, size, and associated uncertainties.

\subsection{Comparison with the L19 CDA} 
\label{sub:comparison_with_the_lee2019_cda}

The L19 CDA is conceptually simpler than the CDA presented here, using the \citet{Silburt2019} UNET architecture to find crater candidates, followed by local and global deduplication. The additional work in this CDA brings in the IR dataset that better resolves the smaller craters (stage 1), includes additional verification of the crater candidates (stage 2), and refines the final catalog by considering the duplicate count and other metrics to produce a more accurate final catalog (stage 3).

The source DTM dataset is the same for this work and L19, and the new CDA also uses THEMIS daytime IR imagery\citep{Edwards2011}. As in L19, empirically obtained uncertainties are about 1 pixel in diameter and location, which scales with the pixel scale of the source image but is lower than 5\% for almost all detectable craters.

In contrast to the L19 CDA, we adjusted the stage 1 network parameters to have higher overall recall and lower precision, while achieving a similar F1 score \citep{Lee2019}. This change allowed us to create a high recall catalog in stage 1 (i.e., find many craters), and remove additional false--positive detections using later post--processing stages.

The increase in the number of neural networks we use in this CDA does not increase training or processing time. The ResUNET architecture trains up to four times faster than the L19 UNET architecture, and improvements in the orthographic projection code \citep{PROJContributors2018} and the template matching code reduce the image processing time by a factor of two. As in L19, the bottleneck in the processing speed is the data generation step, not the neural network steps. Computational limits on the consumer workstation used (64GB RAM, 8 core CPU, 1 NVIDIA 1080Ti GPU) and the size of the input dataset (the DTM data is 6GB uncompressed, the IR is 22GB data uncompressed) limited the throughput of the CDA to around 2,000 images per minute. For the results discussed in the next section, the training and processing time is under 72 hours.


\section{Results and Discussion} 
\label{sec:results_and_discussion}

\subsection{Results}\label{results}

The results presented in this section are from an experiment using the CDA to process the DTM and IR global maps. From these maps, we generated image swatches to systematically cover the planet at sizes from 1.5 degrees to 30 degrees, and additional IR images to cover the planet with image sizes from 0.5 to 1.3 degrees (the IRH dataset). None of these images were used in the training process, and to remove possible overfitting to the \citet{Robbins2012} catalog, we compare our results to the \citet{Salamuniccar2012} catalog. For the imagery datasets we used in this work, this CDA produces a statistically complete catalog \citep{Wang2020}, and performs at expert level when compared to the independent \citet{Salamuniccar2011} catalog and the \citet{Robbins2012} catalog.

Table \ref{table:results} summarizes the global catalogs generated by this CDA compared against \citet{Salamuniccar2012} for craters larger than 3km in diameter. For each of the three input datasets (DTM, IR, and IRH) we processed the entire image set with the stage 1 ResUNET and skipping stage 2 for the columns marked with a star. For the combined dataset, we processed all three datasets the full CDA as described in the methods. 

Smaller craters are far more numerous than larger craters, so the high--resolution IR dataset (IRH) finds almost as many craters as the lower resolution DTM and IR datasets, while only finding craters smaller than 16km in diameter. Each of the stage 1 ResUNETs performs comparably with L19 (measured by the F1 score\citep{Lee2019}), but falls below the performance of \citet{Robbins2012} compared against \citet{Salamuniccar2012}. The combined catalog performs as well or better than \citet{Robbins2012} against \citet{Salamuniccar2012}, and slightly worse when compared directly to \citet{Robbins2012}. The apparent switch in the recall and precision values in the \citet{Robbins2012} column is because \citet{Robbins2012} has more craters than \citet{Salamuniccar2012}, so the precision (number of matching craters/ number of total candidates) is low and the recall (number of matching craters/ number of real craters) is high. Reversing the comparison so that \citet{Robbins2012} is the target catalog would produce an identical F1 score but with swapped \textit{precision} and \textit{recall}.

\begin{table*}[ht]
\begin{tabular}{|c||c|c|c|c|c|c|c|}
\hline
& Combined & DTM$^\star$ & IR$^\star$ & IRH$^\star$ & R12 & S12 & L19 \\
\hline
Craters detected & 57,444 & 52,653 & 49,100 & 45,727 &  79,528 & 62,086 & 58,144\\
Craters matched & 50,264 & 43,786 & 44,735 & 38,840 & 57,905 & 57,915 & 44,407\\
Precision & 87 & 83 & 91 & 84 &  72 & 93 & 76\\
Recall & 80 & 70 & 72 & 62 & 93 & 72 & 71\\
F1 & 84 & 76 & 80 & 72 & 81 & 81 & 73\\
Latitude Error & $ 2 \dpm{1}{2} $ & $ 3 \dpm{2}{2} $ & $ 2 \dpm{1}{2} $ & $ 2 \dpm{1}{2} $ & $ 2 \dpm{1}{2} $ & $ 2 \dpm{1}{2} $ & $ 3 \dpm{2}{3} $ \\ 
Longitude Error & $ 3 \dpm{2}{2} $ & $ 3 \dpm{2}{3} $ & $ 3 \dpm{1}{2} $ & $ 3 \dpm{2}{2} $ & $ 2 \dpm{1}{2} $ & $ 2 \dpm{1}{2} $ & $ 4 \dpm{3}{5} $ \\
Diameter Error & $ 4 \dpm{2}{3} $ & $ 4 \dpm{2}{3} $ & $ 4 \dpm{2}{3} $ & $ 3 \dpm{2}{3} $ & $ 3 \dpm{2}{3} $ & $ 3 \dpm{2}{3} $ & $ 7 \dpm{4}{6} $ \\ 
\hline
\end{tabular}
\caption{\label{table:results} Comparison against the \citet{Salamuniccar2012} catalog. Craters detected and matched are given as number of craters. All other numbers are given in percentages, with errors quoted relative to the minimum diameter used in the comparison between the generated catalog and target catalog. The DTM, IR, and IRH columns show the results from processing only DTM, only IR above 1.3 degree image swatches, and only IR below 1.3 degree image swatches (`high' resolution). The combined column shows all 3 datasets combined by the CDA. The R12 column compares the \citet{Robbins2012} catalog against \citet{Salamuniccar2012}. The S12 column compared \citet{Salamuniccar2012} against \citet{Robbins2012}, reversing the R12 comparison.}
\end{table*}

\begin{figure}
\centering
\includegraphics[scale=0.75]{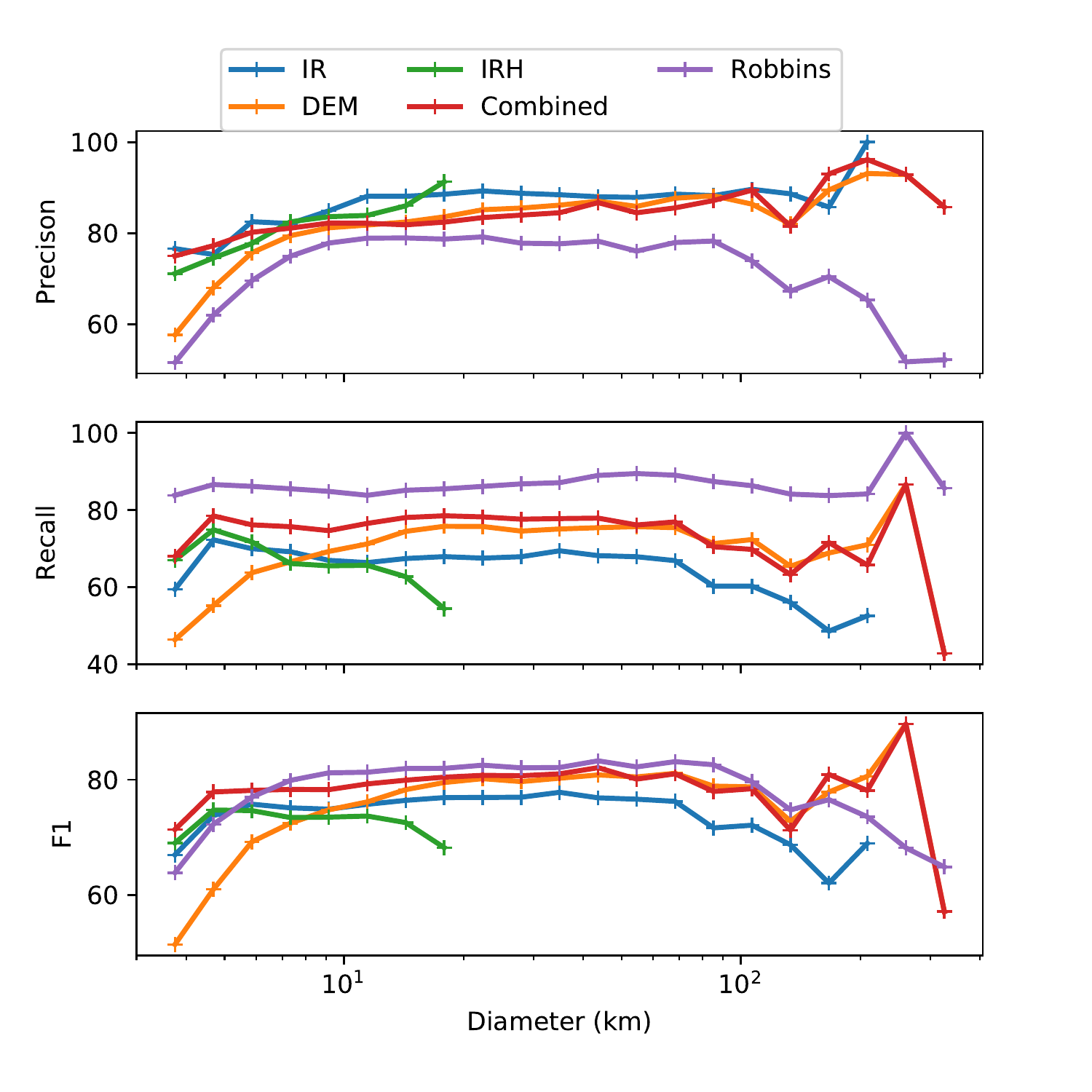}
\caption{\label{fig:prf} Precision, Recall, and F1 score as a function of crater diameter for the individual CDAs for IRH, IR, DTM, and for the combined CDA. Each datapoint is calculated using craters with diameters within 20\% of the marked datapoint, and all values are given in percentage units. The comparable test using \citet{Robbins2012} is included for reference.}
\end{figure}

Figure \ref{fig:prf} shows the precision, recall, and F1 scores for each of the columns in Table \ref{table:results}, grouped by the crater diameter. For each data point, we calculate the diagnostics considering all craters that lie within 20\% of each marker. The drop--off in DTM \textit{recall} at smaller diameters is the main reason the DTM catalog performs more poorly than the IR catalog overall --- smaller crater are far more numerous and have a significant effect on these metrics. The drop--off in the IRH catalog with increasing diameter is worse than the IR catalog because of the lack of larger--scale images to contribute to the catalog.

\begin{figure}
\centering
\includegraphics[scale=0.75]{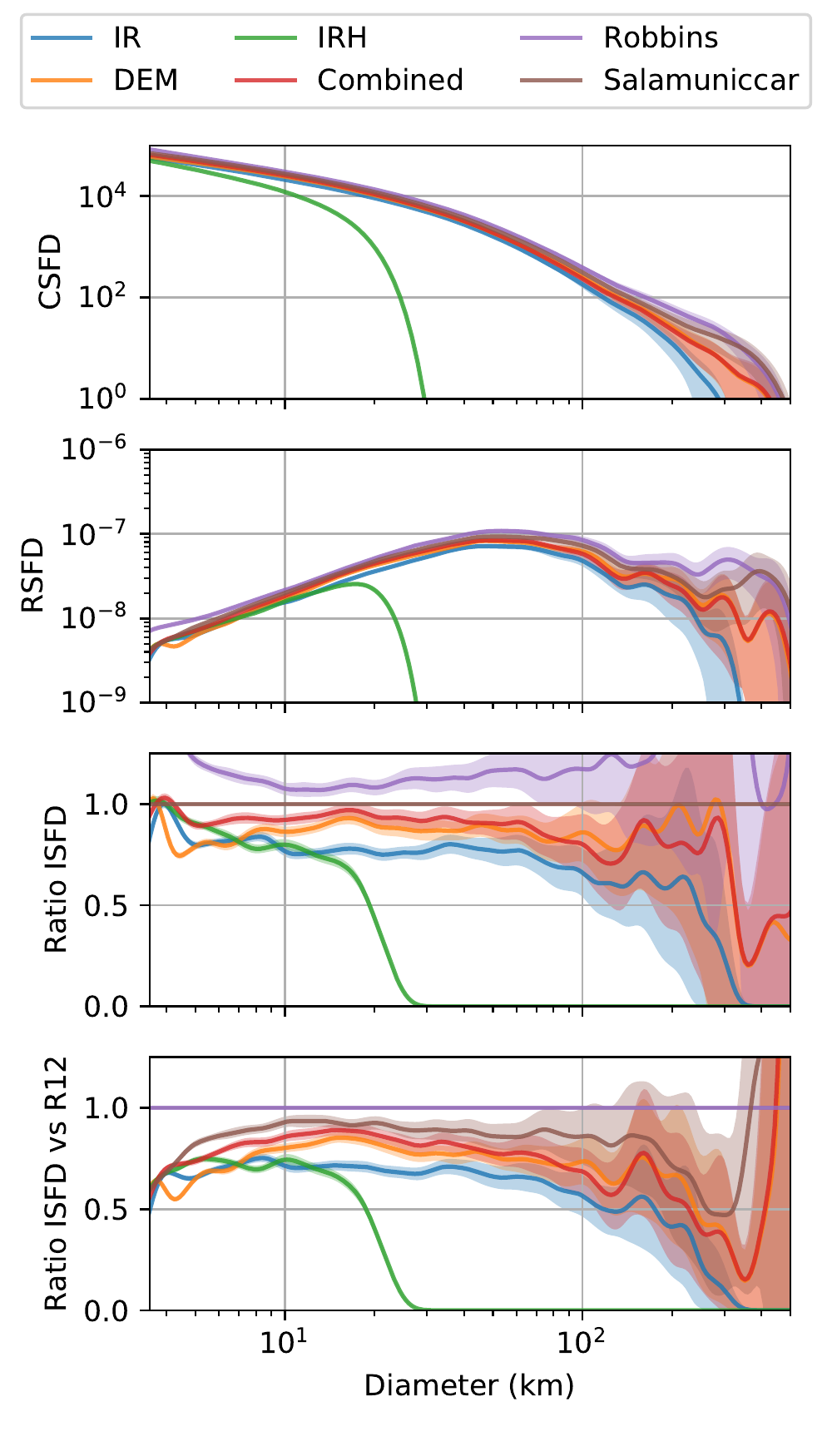}
    \caption{\label{fig:sfd} SFDs for the crater catalogs generated from the IR, DTM, and combined catalogs, as well as \citet{Robbins2012} and \citet{Salamuniccar2012} Mars catalogs. (top) Cumulative SFD, (middle) Relative SFD, (bottom) ratio of Incremental SFD to the \citet{Robbins2012} ISFD. SFDs are calculated following \citet{Robbins2018} using the recommended bandwidth of 0.1D and a Gaussian kernel.}
\end{figure}

Figure \ref{fig:sfd} shows crater metrics recommended by \citet{Robbins2018} for comparing crater catalogs. In particular, we use the smoothed `Size--Frequency Distribution` (SFD) algorithm to calculate the distribution of crater sizes instead of the histogram method used in earlier work \citep{Arvidson1974}. In contrast to the recall and precision data in Figure \ref{fig:prf} and Table \ref{table:results}, the SFD calculation only considers crater numbers (per diameter) and does not match craters between catalogs first, so two catalogs with similar SFDs only match in statistical population count, not in specific crater identifications. Figure \ref{fig:sfd} shows three versions of the SFD following \citet{Robbins2018}. The cumulative SFD (CSFD, similar to a raw crater count), the relative SFD (RSFD, similar to the metric suggested by the Crater Analysis and Techniques Working Group \citep{Arvidson1974}), and the ratio of incremental SFDs (ISFD) to compare the relative crater counts between catalogs. In this Figure, the ISFDs are compared to the \citet{Robbins2012} catalog.

\begin{figure}
\centering
\includegraphics[width=1.0\textwidth]{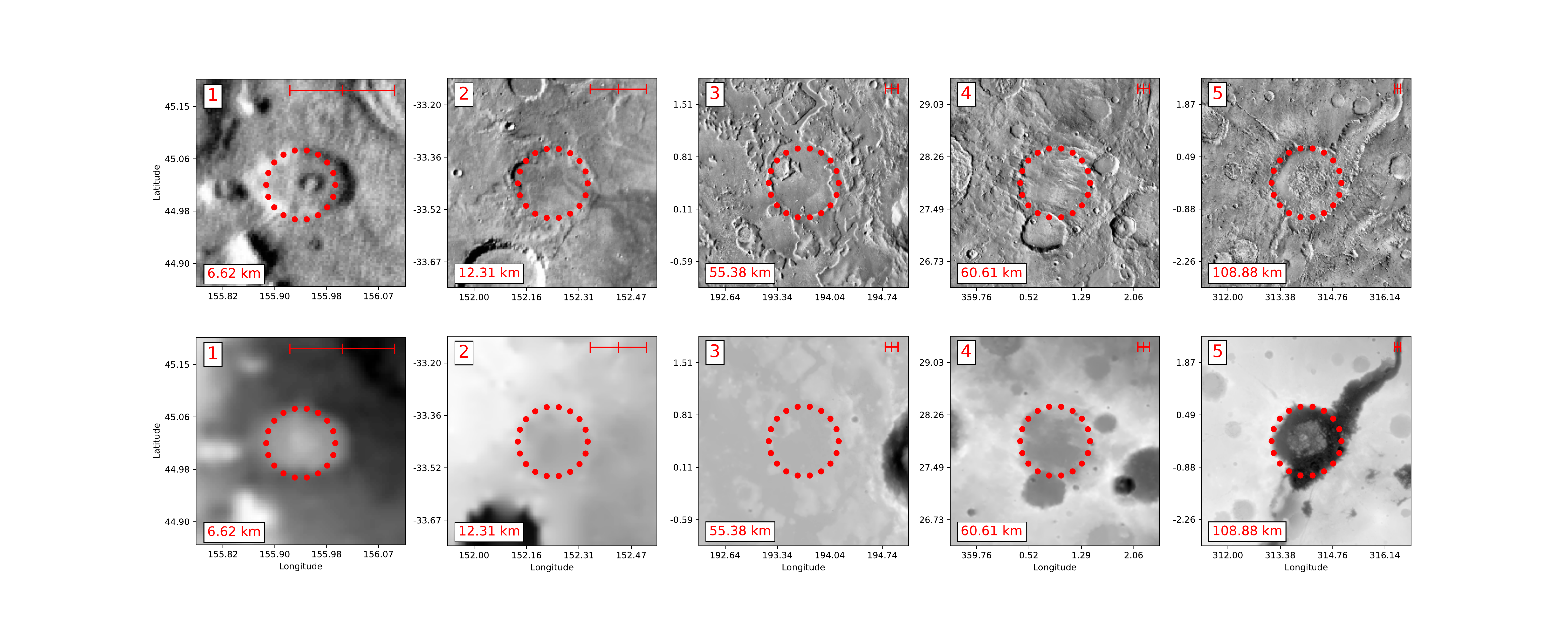}
\caption{\label{fig:fp} A sample of 5 `new' crater candidates found by the CDA. Each image is a randomly chosen crater from the catalog with diameter in the range (left to right) 6--10km, 10--40km, 40--60km, 60--100km, 100--400km. The THEMIS IR (top) and MOLA/HRSC DTM (bottom) images are shown for each crater, along with the CDA measured diameter, and a 10km scale bar (marked at 0,5,10km) for reference. Crater 1,2, and 5 are included in the \citet{Robbins2012} catalog but not in \citet{Salamuniccar2012}, and would not be considered `new' in a comparison between the CDA and \citet{Robbins2012}.}
\end{figure}

\begin{figure}
\centering
\includegraphics[width=1.0\textwidth]{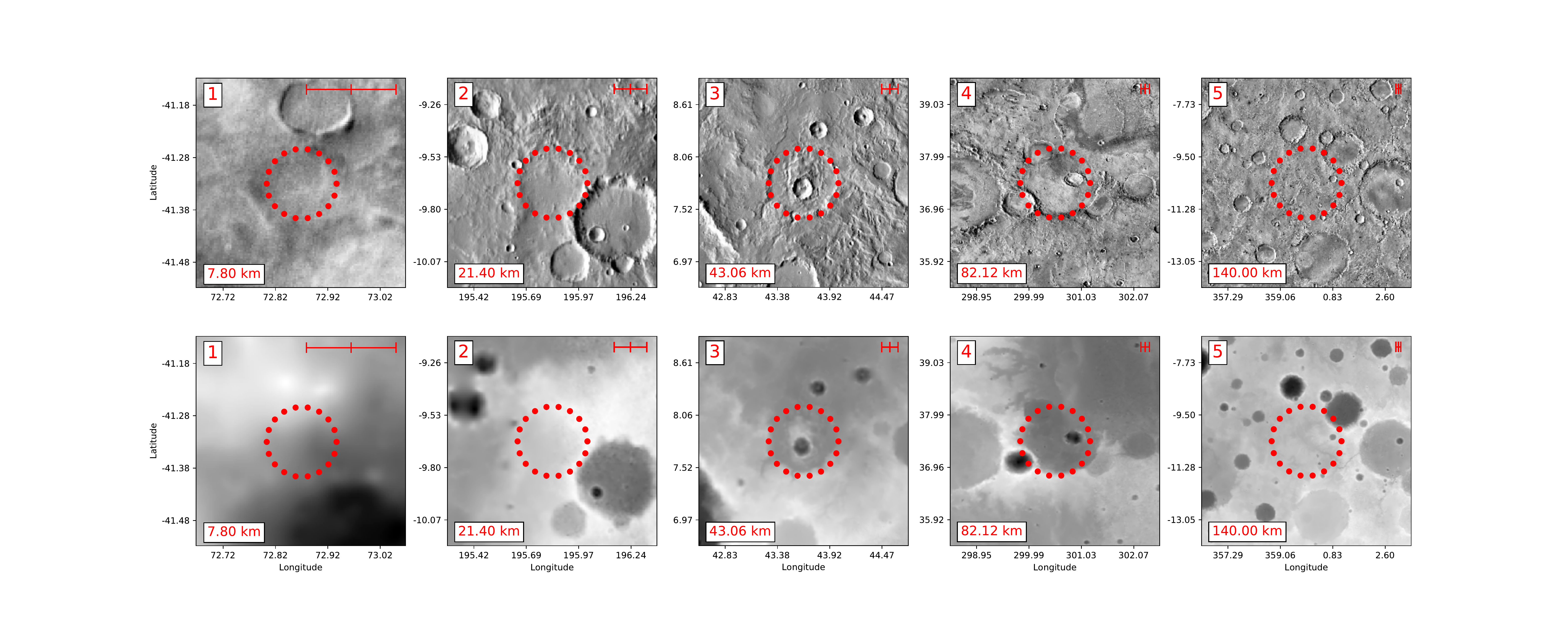}
\caption{\label{fig:fn} A sample of 5 craters missed by the CDA. The Figure is constructed identically to Figure \ref{fig:fp} but using the craters missed by the CDA and included in \citet{Salamuniccar2012}. The craters at the center of images 1 and 5 are not included in \citet{Robbins2012} but are in \citet{Salamuniccar2012}. The CDA finds the crater in images 2 and 3 in the DTM, but it is rejected with a low probability from the classifier network and no IR detections. The crater in image 4 was not found by the CDA.}
\end{figure}

Following \citet{Lee2019}, we show a sample of `new' crater candidates in  Figure \ref{fig:fp} and a sample of missed craters in Figure \ref{fig:fn}. We generate each image as in L19 with the crater centered in the image with 1 crater diameter on each side. In the metrics calculated for Table \ref{table:results} the `new' craters (Figure \ref{fig:fp}) are considered false positives and reduce the precision, while the missed craters (Figure \ref{fig:fn}) are considered false negatives and reduce the recall.

\subsection{Discussion}\label{discussion}

The CDA developed here performs at expert level down to 3km diameter craters. This new CDA has an overall (F1 score) performance on par with the \citet{Robbins2012} catalog in comparison to \citet{Salamuniccar2012}, or vice versa. The CDA is 10\% more precise and has a 5\% higher recall than the L19 CDA compared against \citet{Salamuniccar2012}, with similar numbers when compared to \citet{Robbins2012}.

We use multiple stages in the new CDA to find many crater candidates (stage 1) and refine the candidates into a crater catalog (stage 2 and 3). In contrast to the L19, our stage 1 ResUNET is more `liberal' at identifying craters, potentially finding more false--positive craters and allowing the later stages to remove some of these false--positive detections.

The later stages of the CDA help to merge disparate catalogs from the IR and DTM CDAs. We combined the candidates from the IR,IRH, and DTM catalogs in the `combined' catalog, increasing the total crater count, recal, and F1 score. The precision of the combined catalog does drop compared to the IR catalog, but the combined catalog includes more than 6,000 new matching craters for the 2,000 extra non--matching candidates.

The THEMIS IR dataset extended the range of detectable craters below the native resolution of the MOLA/HRSC dataset\citep{Fergason2018}. We limited the analysis in L19 to about 3.8km diameter, or 8 pixels at the MOLA scale of 463m/pixel. The 100m/pixel scale of the THEMIS dataset should allow a lower diameter of 1km. We used a conservative limit of 3km diameter for this work because the stage 2 network relies on DTM data to work, even if the stage 1 input data is higher resolution.

We find that the position and diameter `errors' in matching craters are similar across the different datasets compared to \citet{Salamuniccar2012}, and are similar to the L19 results. For most metrics in Table \ref{table:results} the median difference between our catalog and the \citet{Salamuniccar2012} catalog is 2\% relative to each crater's diameter, or equivalently, less than 1 pixel. If we include the uncertainties calculated by our CDA, matching craters from \citet{Salamuniccar2012} are within 1$\sigma$ for 76\% of the craters in our catalog.

In contrast, we find most of the unique craters in each catalog are genuinely different from craters in the other catalog, and not near--matches. For false--positive (`new') craters and for false--negative (missed) craters, the nearest crater that meets the size criterion (the crater diameters are within 25\% of each other) is five crater widths away on average. Fewer than 10\% of either population are within one crater width of a crater in the other catalog.

False--positive (`new') crater candidates include a few plausible craters that may be missing from the \textit{ground--truth} catalog, but many more are implausible candidates such as curved land formations or ill--resolved depressions (Figure \ref{fig:fp}). Nevertheless, the false--positive rate has been reduced by a factor of 2 from L19, with much of the improvement due to the later stages filtering out anomalous small crater identifications. As in L19, The false--positive list includes a small number of volcanic paterae that are not explicitly excluded from the CDA.

False--negative craters include disagreements in location or size but are dominated by genuine misses of degraded and shallow craters (Figure \ref{fig:fn}). The CDA is twice as likely to miss a highly degraded crater (\texttt{DEGRADATION\_STATE} of 1 in \citet{Robbins2012}) compared to the next most degraded state. In addition, each 100m in crater depth increases the chance of detection by 30\%, and each 1\% decrease in eccentricity increases the chance of detection by 3\%. In all cases, the diameter of the crater does not significantly affect these results.

\subsubsection{Limitations}
Notwithstanding the improvements to the CDA, we have added additional complexity to the algorithm and additional requirements before the CDA can be used on a new planetary body. The L19 CDA (and \citet{Silburt2019} CDA) did not require any sample of craters from the body before the CDA could be used --- the UNET was tested \textit{without training} on Mercury\citep{Silburt2019} and Pluto\citep{Lee2018a} Our new CDA benefits, in part, from the stage 2 classifier network that requires a sample of \textit{ground--truth} data, and the stage 3 filter that requires independent validation of a sample of the CDA candidates.

In the experiment described above, we used the same \textit{ground--truth} catalog to train the stage 2 network and validate the CDA candidates in stage 3, but in principle the workflow can be adjusted to accommodate new data. Stage 2 requires a sample of ground--truth craters that could be manually measured from a region of a new planetary body. Stage 3 requires validation of candidates proposed by the CDA, which can be quickly validated by humans. The CDA can work without stage 2, losing a few percent in precision and accuracy, and performs about as well as L19 without stage 3.

\subsubsection{Future Improvements} 
We have improved the L19 CDA to be competitive with human--generated catalogs. Improvements beyond this limit are difficult using human generated \textit{ground--truth} datasets without more complex neural networks. It is also not clear that better agreement between catalogs should be targeted in future work \citep{Robbins2014}. Instead, refinements in the pre--processing stages, and calculating additional crater metrics would likely be more productive additions to the algorithm than increasing naive crater matching performance.

For example, the pre--processing stages could be improved to expose shallow craters, either by smoothing to varying spatial resolutions \citep{Stepinski2009} or adjusting contrast in variable terrain to highlight shallow features \citep{Lee2019}. The CDA could also be used to process smaller scale localized images as part of ongoing \textit{mission operations} to provide value--added data to the images of the surface \citep[e.g.,][]{Wronkiewicz2018,Lee2018a}.

Two metrics are relatively straightforward to extract from the DTM and IR datasets, ellipticity and depth. \citet{Lee2018a} and \citet{Ali-Dib2019} showed that a MaskRCNN network\citep{He2017} can extract crater--rim points that can then be fit to ellipses \citet[e.g.][]{Robbins2018Ellipse} to provide ellipticity and orientation data for the craters (for use in secondary crater mapping \citep{Naegeli2019}). \citet{Stepinski2009} attempted to determine crater depth automatically from terrain data, and \citet{Robbins2018a} discusses using IR imagery for similar purposes. \citet{Robbins2012} provides ground--truth data for both of these metrics, so networks could be trained to replicate their performance.

\section{Conclusions} 
\label{sec:conclusions}
This paper develops a CDA to find near--circular craters in Mars DTM and IR imagery. The CDA improves on the \citet{Lee2019} CDA by using IR data to extend the lower diameter limit to below 3km and by using new post--processing stages to verify and refine the crater catalogs generated from the imagery data.

The performance of the CDA measured against \citet{Salamuniccar2012} matches the performance of catalogs generated by expert human classifiers \citep{Robbins2012} while working hundreds of times faster than typical human classification speeds. By design, the CDA incorporates multiple crater detection networks that perform with similar skill to the \citet{Lee2019} algorithm, combined with a post--processing network designed to improve the precision of the algorithm to reduce the occurrence of false crater identifications.

The new CDA performs worst on shallow crater and degraded craters, as does L19. Shallow crater detection was improved using IR imagery, but we did not include additional pre--processing of images to expose shallow features in the DTM\citep[e.g., ][]{Stepinski2009b}. 

The post--processing stages (2 and 3) improve precision of the CDA by removing false--positive candidates without sacrificing recall by removing many true craters. These stages could be used to improve results from other CDA \citep[e.g.][]{Wronkiewicz2018,Ali-Dib2019}, or be retrained to refine catalogs of other features on the surface of planetary bodies\citep[e.g.,][]{Aye2019,Bickel2019}.


\section{Acknowledgments}

We thank the three anonymous reviewers and the journal editors for their constructive comments that helped improve the clarity of the paper. Computations were performed on clusters and GPU enabled workstations at the University of Toronto. J. H. was funded by an Undergraduate Student Research Award from the Natural Sciences and Engineering Research Council of Canada during summer 2019.

\section{Computer Code Availability}

The crater detection software, named DeepMars2 (to distinguish from the L19 algorithm named DeepMars) is available with an MIT licence from the Dataverse repository for this manuscript (\url{ https://doi.org/10.5683/SP2/CFUNII}). The CDA software is written in Python and requires a standard array of Python libraries to operate. The code will work on workstations with at least 16GB of available memory, but will benefit from additional RAM to open the large input data maps, and an NVIDIA GPU (NVIDIA 1080Ti or larger memory) to accelerate learning and prediction.


\section{Bibliography}
\begin{thebibliography}{44}
\expandafter\ifx\csname natexlab\endcsname\relax\def\natexlab#1{#1}\fi
\providecommand{\url}[1]{\texttt{#1}}
\providecommand{\href}[2]{#2}
\providecommand{\path}[1]{#1}
\providecommand{\DOIprefix}{doi:}
\providecommand{\ArXivprefix}{arXiv:}
\providecommand{\URLprefix}{URL: }
\providecommand{\Pubmedprefix}{pmid:}
\providecommand{\doi}[1]{\href{http://dx.doi.org/#1}{\path{#1}}}
\providecommand{\Pubmed}[1]{\href{pmid:#1}{\path{#1}}}
\providecommand{\bibinfo}[2]{#2}
\ifx\xfnm\relax \def\xfnm[#1]{\unskip,\space#1}\fi
\bibitem[{Ali-Dib et~al.(2019)Ali-Dib, Menou, Zhu, Hammond and
  Jackson}]{Ali-Dib2019}
\bibinfo{author}{Ali-Dib, M.}, \bibinfo{author}{Menou, K.},
  \bibinfo{author}{Zhu, C.}, \bibinfo{author}{Hammond, N.},
  \bibinfo{author}{Jackson, A.P.}, \bibinfo{year}{2019}.
\newblock \bibinfo{title}{{Automated crater shape retrieval using
  weakly-supervised deep learning}}.
\newblock \bibinfo{journal}{arXiv preprint} \URLprefix
  \url{http://arxiv.org/abs/1906.08826},
  \href{http://arxiv.org/abs/1906.08826}{\tt arXiv:1906.08826}.
\bibitem[{Arvidson(1974)}]{Arvidson1974}
\bibinfo{author}{Arvidson, R.E.}, \bibinfo{year}{1974}.
\newblock \bibinfo{title}{{Morphologic classification of Martian craters and
  some implications}}.
\newblock \bibinfo{journal}{Icarus} \bibinfo{volume}{22},
  \bibinfo{pages}{264--271}.
\newblock \DOIprefix\doi{10.1016/0019-1035(74)90176-6}.
\bibitem[{Aye et~al.(2019)Aye, Schwamb, Portyankina, Hansen, McMaster, Miller,
  Carstensen, Snyder, Parrish, Lynn, Mai, Miller, Simpson and Smith}]{Aye2019}
\bibinfo{author}{Aye, K.M.}, \bibinfo{author}{Schwamb, M.E.},
  \bibinfo{author}{Portyankina, G.}, \bibinfo{author}{Hansen, C.J.},
  \bibinfo{author}{McMaster, A.}, \bibinfo{author}{Miller, G.R.},
  \bibinfo{author}{Carstensen, B.}, \bibinfo{author}{Snyder, C.},
  \bibinfo{author}{Parrish, M.}, \bibinfo{author}{Lynn, S.},
  \bibinfo{author}{Mai, C.}, \bibinfo{author}{Miller, D.},
  \bibinfo{author}{Simpson, R.J.}, \bibinfo{author}{Smith, A.M.},
  \bibinfo{year}{2019}.
\newblock \bibinfo{title}{{Planet Four: Probing springtime winds on Mars by
  mapping the southern polar CO2 jet deposits}}.
\newblock \bibinfo{journal}{Icarus} \bibinfo{volume}{319},
  \bibinfo{pages}{558--598}.
\newblock \URLprefix \url{https://doi.org/10.1016/j.icarus.2018.08.018},
  \DOIprefix\doi{10.1016/j.icarus.2018.08.018}.
\bibitem[{Barlow(1988)}]{Barlow1988}
\bibinfo{author}{Barlow, N.G.}, \bibinfo{year}{1988}.
\newblock \bibinfo{title}{{Crater size-frequency distributions and a revised
  Martian relative chronology}}.
\newblock \bibinfo{journal}{Icarus} \bibinfo{volume}{75},
  \bibinfo{pages}{285--305}.
\newblock \DOIprefix\doi{10.1016/0019-1035(88)90006-1}.
\bibitem[{Barlow and Perez(2003)}]{Barlow2003}
\bibinfo{author}{Barlow, N.G.}, \bibinfo{author}{Perez, C.B.},
  \bibinfo{year}{2003}.
\newblock \bibinfo{title}{{Martian impact crater ejecta morphologies as
  indicators of the distribution of subsurface volatiles}}.
\newblock \bibinfo{journal}{Journal of Geophysical Research}
  \bibinfo{volume}{108}, \bibinfo{pages}{5085}.
\newblock \URLprefix \url{http://doi.wiley.com/10.1029/2002JE002036},
  \DOIprefix\doi{10.1029/2002JE002036}.
\bibitem[{Bickel et~al.(2019)Bickel, Lanaras, Manconi, Loew and
  Mall}]{Bickel2019}
\bibinfo{author}{Bickel, V.T.}, \bibinfo{author}{Lanaras, C.},
  \bibinfo{author}{Manconi, A.}, \bibinfo{author}{Loew, S.},
  \bibinfo{author}{Mall, U.}, \bibinfo{year}{2019}.
\newblock \bibinfo{title}{{Automated Detection of Lunar Rockfalls Using a
  Convolutional Neural Network}}.
\newblock \bibinfo{journal}{IEEE Transactions on Geoscience and Remote Sensing}
  \bibinfo{volume}{57}, \bibinfo{pages}{3501--3511}.
\newblock \DOIprefix\doi{10.1109/TGRS.2018.2885280}.
\bibitem[{Bue and Stepinski(2006)}]{Bue2006}
\bibinfo{author}{Bue, B.D.}, \bibinfo{author}{Stepinski, T.F.},
  \bibinfo{year}{2006}.
\newblock \bibinfo{title}{{Automated classification of landforms on Mars}}.
\newblock \bibinfo{journal}{Computers and Geosciences} \bibinfo{volume}{32},
  \bibinfo{pages}{604--614}.
\newblock \DOIprefix\doi{10.1016/j.cageo.2005.09.004}.
\bibitem[{Chollet and Others(2015)}]{Chollet2015}
\bibinfo{author}{Chollet, F.}, \bibinfo{author}{Others}, \bibinfo{year}{2015}.
\newblock \bibinfo{title}{{Keras}}.
\newblock \URLprefix \url{https://github.com/fchollet/keras}.
\bibitem[{Cintala et~al.(1976)Cintala, Head and Mutch}]{Cintala1976}
\bibinfo{author}{Cintala, M.J.}, \bibinfo{author}{Head, J.W.},
  \bibinfo{author}{Mutch, T.A.}, \bibinfo{year}{1976}.
\newblock \bibinfo{title}{{Martian crater depth/diameter relationships :
  Comparison with the Moon and Mercury}}, in: \bibinfo{booktitle}{Proc. Lunar.
  Sci. Conf. 7th}, pp. \bibinfo{pages}{3575--3587}.
\bibitem[{Di et~al.(2014)Di, Li, Yue, Sun and Liu}]{Di2014}
\bibinfo{author}{Di, K.}, \bibinfo{author}{Li, W.}, \bibinfo{author}{Yue, Z.},
  \bibinfo{author}{Sun, Y.}, \bibinfo{author}{Liu, Y.}, \bibinfo{year}{2014}.
\newblock \bibinfo{title}{{A machine learning approach to crater detection from
  topographic data}}.
\newblock \bibinfo{journal}{Advances in Space Research} \bibinfo{volume}{54},
  \bibinfo{pages}{2419--2429}.
\newblock \URLprefix \url{http://dx.doi.org/10.1016/j.asr.2014.08.018},
  \DOIprefix\doi{10.1016/j.asr.2014.08.018}.
\bibitem[{Edwards et~al.(2011)Edwards, Nowicki, Christensen, Hill, Gorelick and
  Murray}]{Edwards2011}
\bibinfo{author}{Edwards, C.S.}, \bibinfo{author}{Nowicki, K.J.},
  \bibinfo{author}{Christensen, P.R.}, \bibinfo{author}{Hill, J.},
  \bibinfo{author}{Gorelick, N.}, \bibinfo{author}{Murray, K.},
  \bibinfo{year}{2011}.
\newblock \bibinfo{title}{{Mosaicking of global planetary image datasets: 1.
  Techniques and data processing for Thermal Emission Imaging System (THEMIS)
  multi-spectral data}}.
\newblock \bibinfo{journal}{Journal of Geophysical Research E: Planets}
  \bibinfo{volume}{116}, \bibinfo{pages}{1--21}.
\newblock \DOIprefix\doi{10.1029/2010JE003755}.
\bibitem[{Fergason et~al.(2018)Fergason, Hare and Laura}]{Fergason2018}
\bibinfo{author}{Fergason, R.}, \bibinfo{author}{Hare, T.},
  \bibinfo{author}{Laura, J.}, \bibinfo{year}{2018}.
\newblock \bibinfo{title}{{HRSC and MOLA Blended Digital Elevation Model at
  200m v2}}.
\newblock \URLprefix \url{http://bit.ly/HRSC{\_}MOLA{\_}Blend{\_}v0}.
\bibitem[{Goodfellow et~al.(2016)Goodfellow, Bengio and
  Courville}]{Goodfellow2016}
\bibinfo{author}{Goodfellow, I.}, \bibinfo{author}{Bengio, Y.},
  \bibinfo{author}{Courville, A.}, \bibinfo{year}{2016}.
\newblock \bibinfo{title}{{Deep Learning}}.
\newblock \bibinfo{publisher}{MIT Press}.
\bibitem[{He et~al.(2017)He, Gkioxari, Dollar and Girshick}]{He2017}
\bibinfo{author}{He, K.}, \bibinfo{author}{Gkioxari, G.},
  \bibinfo{author}{Dollar, P.}, \bibinfo{author}{Girshick, R.},
  \bibinfo{year}{2017}.
\newblock \bibinfo{title}{{Mask R-CNN}}.
\newblock \bibinfo{journal}{Proceedings of the IEEE International Conference on
  Computer Vision} \bibinfo{volume}{2017-Octob}, \bibinfo{pages}{2980--2988}.
\newblock \DOIprefix\doi{10.1109/ICCV.2017.322},
  \href{http://arxiv.org/abs/1703.06870}{\tt arXiv:1703.06870}.
\bibitem[{Kinczyk et~al.(2020)Kinczyk, Prockter, Byrne, Susorney and
  Chapman}]{Kinczyk2020}
\bibinfo{author}{Kinczyk, M.J.}, \bibinfo{author}{Prockter, L.M.},
  \bibinfo{author}{Byrne, P.K.}, \bibinfo{author}{Susorney, H.C.},
  \bibinfo{author}{Chapman, C.R.}, \bibinfo{year}{2020}.
\newblock \bibinfo{title}{{A morphological evaluation of crater degradation on
  Mercury: Revisiting crater classification with MESSENGER data}}.
\newblock \bibinfo{journal}{Icarus} \bibinfo{volume}{341},
  \bibinfo{pages}{113637}.
\newblock \URLprefix \url{https://doi.org/10.1016/j.icarus.2020.113637},
  \DOIprefix\doi{10.1016/j.icarus.2020.113637}.
\bibitem[{Kr{\o}gli and Dypvik(2010)}]{Krogli2010}
\bibinfo{author}{Kr{\o}gli, S.O.}, \bibinfo{author}{Dypvik, H.},
  \bibinfo{year}{2010}.
\newblock \bibinfo{title}{{Automatic detection of circular outlines in regional
  gravity and aeromagnetic data in the search for impact structure
  candidates}}.
\newblock \bibinfo{journal}{Computers and Geosciences} \bibinfo{volume}{36},
  \bibinfo{pages}{477--488}.
\newblock \URLprefix \url{http://dx.doi.org/10.1016/j.cageo.2009.07.010},
  \DOIprefix\doi{10.1016/j.cageo.2009.07.010}.
\bibitem[{Lee(2018)}]{Lee2018a}
\bibinfo{author}{Lee, C.}, \bibinfo{year}{2018}.
\newblock \bibinfo{title}{{Martian Crater Identification Using Deep Learning}},
  in: \bibinfo{booktitle}{American Geophysical Union Fall Meeting}, pp.
  \bibinfo{pages}{P41D--3768}.
\bibitem[{Lee(2019)}]{Lee2019}
\bibinfo{author}{Lee, C.}, \bibinfo{year}{2019}.
\newblock \bibinfo{title}{{Automated Crater Detection on Mars using Deep
  Learning}}.
\newblock \bibinfo{journal}{Planetary and Space Science} \bibinfo{volume}{170}, \DOIprefix\doi{10.1016/j.pss.2019.03.008}.
\bibitem[{Milletari et~al.(2016)Milletari, Navab and Ahmadi}]{Milletari2016}
\bibinfo{author}{Milletari, F.}, \bibinfo{author}{Navab, N.},
  \bibinfo{author}{Ahmadi, S.A.}, \bibinfo{year}{2016}.
\newblock \bibinfo{title}{{V-Net: Fully convolutional neural networks for
  volumetric medical image segmentation}}.
\newblock \bibinfo{journal}{Proceedings - 2016 4th International Conference on
  3D Vision, 3DV 2016} ,
  \bibinfo{pages}{565--571}\DOIprefix\doi{10.1109/3DV.2016.79},
  \href{http://arxiv.org/abs/1606.04797}{\tt arXiv:1606.04797}.
\bibitem[{Naegeli and Laura(2019)}]{Naegeli2019}
\bibinfo{author}{Naegeli, T.J.}, \bibinfo{author}{Laura, J.},
  \bibinfo{year}{2019}.
\newblock \bibinfo{title}{{Back-projecting secondary craters using a cone of
  uncertainty}}.
\newblock \bibinfo{journal}{Computers and Geosciences} \bibinfo{volume}{123},
  \bibinfo{pages}{1--9}.
\newblock \URLprefix \url{https://doi.org/10.1016/j.cageo.2018.10.011},
  \DOIprefix\doi{10.1016/j.cageo.2018.10.011}.
\bibitem[{Palafox et~al.(2017)Palafox, Hamilton, Scheidt and
  Alvarez}]{Palafox2017}
\bibinfo{author}{Palafox, L.F.}, \bibinfo{author}{Hamilton, C.W.},
  \bibinfo{author}{Scheidt, S.P.}, \bibinfo{author}{Alvarez, A.M.},
  \bibinfo{year}{2017}.
\newblock \bibinfo{title}{{Automated detection of geological landforms on Mars
  using Convolutional Neural Networks}}.
\newblock \bibinfo{journal}{Computers and Geosciences} \bibinfo{volume}{101},
  \bibinfo{pages}{48--56}.
\newblock \URLprefix \url{http://dx.doi.org/10.1016/j.cageo.2016.12.015},
  \DOIprefix\doi{10.1016/j.cageo.2016.12.015}.
\bibitem[{Palucis et~al.(2020)Palucis, Jasper, Garczynski and
  Dietrich}]{Palucis2020}
\bibinfo{author}{Palucis, M.C.}, \bibinfo{author}{Jasper, J.},
  \bibinfo{author}{Garczynski, B.}, \bibinfo{author}{Dietrich, W.E.},
  \bibinfo{year}{2020}.
\newblock \bibinfo{title}{{Quantitative assessment of uncertainties in modeled
  crater retention ages on Mars}}.
\newblock \bibinfo{journal}{Icarus} \bibinfo{volume}{341},
  \bibinfo{pages}{113623}.
\newblock \URLprefix \url{https://doi.org/10.1016/j.icarus.2020.113623},
  \DOIprefix\doi{10.1016/j.icarus.2020.113623}.
\bibitem[{Pedregosa et~al.(2011)Pedregosa, Varoquaux, Gramfort, Michel,
  Thirion, Grisel, Blondel, Prettenhofer, Weiss, Dubourg, Vanderplas, Passos,
  Cournapeau, Brucher, Perrot and Duchesnay}]{Pedregosa}
\bibinfo{author}{Pedregosa, F.}, \bibinfo{author}{Varoquaux, G.},
  \bibinfo{author}{Gramfort, A.}, \bibinfo{author}{Michel, V.},
  \bibinfo{author}{Thirion, B.}, \bibinfo{author}{Grisel, O.},
  \bibinfo{author}{Blondel, M.}, \bibinfo{author}{Prettenhofer, P.},
  \bibinfo{author}{Weiss, R.}, \bibinfo{author}{Dubourg, V.},
  \bibinfo{author}{Vanderplas, J.}, \bibinfo{author}{Passos, A.},
  \bibinfo{author}{Cournapeau, D.}, \bibinfo{author}{Brucher, M.},
  \bibinfo{author}{Perrot, M.}, \bibinfo{author}{Duchesnay, E.},
  \bibinfo{year}{2011}.
\newblock \bibinfo{title}{{Scikit-learn: Machine Learning in Python}}.
\newblock \bibinfo{journal}{Journal of Machine Learning Research}
  \bibinfo{volume}{12}, \bibinfo{pages}{2825--2830}.
\bibitem[{Pedrosa et~al.(2017)Pedrosa, de~Azevedo, da~Silva and
  Dias}]{Pedrosa2017}
\bibinfo{author}{Pedrosa, M.M.}, \bibinfo{author}{de~Azevedo, S.C.},
  \bibinfo{author}{da~Silva, E.A.}, \bibinfo{author}{Dias, M.A.},
  \bibinfo{year}{2017}.
\newblock \bibinfo{title}{{Improved automatic impact crater detection on Mars
  based on morphological image processing and template matching}}.
\newblock \bibinfo{journal}{Geomatics, Natural Hazards and Risk}
  \bibinfo{volume}{8}, \bibinfo{pages}{1306--1319}.
\newblock \URLprefix \url{https://doi.org/10.1080/19475705.2017.1327463},
  \DOIprefix\doi{10.1080/19475705.2017.1327463}.
\bibitem[{{PROJ Contributors}(2018)}]{PROJContributors2018}
\bibinfo{author}{{PROJ Contributors}}, \bibinfo{year}{2018}.
\newblock \bibinfo{title}{{PROJ Coordinate Transformation Software Library}}.
\bibitem[{Redmon et~al.(2016)Redmon, Divvala, Girshick and
  Farhadi}]{Redmon2016}
\bibinfo{author}{Redmon, J.}, \bibinfo{author}{Divvala, S.},
  \bibinfo{author}{Girshick, R.}, \bibinfo{author}{Farhadi, A.},
  \bibinfo{year}{2016}.
\newblock \bibinfo{title}{{You only look once: Unified, real-time object
  detection}}.
\newblock \bibinfo{journal}{Proceedings of the IEEE Computer Society Conference
  on Computer Vision and Pattern Recognition} \bibinfo{volume}{2016-Decem},
  \bibinfo{pages}{779--788}.
\newblock \DOIprefix\doi{10.1109/CVPR.2016.91},
  \href{http://arxiv.org/abs/1506.02640}{\tt arXiv:1506.02640}.
\bibitem[{Robbins(2018)}]{Robbins2018Ellipse}
\bibinfo{author}{Robbins, S.J.}, \bibinfo{year}{2018}.
\newblock \bibinfo{title}{{A potpourri of related crater cataloging: From
  fitting ellipses to new basemaps to crater production functions}}.
\newblock \URLprefix
  \url{papers2://publication/uuid/BFDC7D14-EB5B-46FF-A6B7-964CD98C7413},
  \DOIprefix\doi{10.1023/A}.
\bibitem[{Robbins et~al.(2014)Robbins, Antonenko, Kirchoff, Chapman, Fassett,
  Herrick, Singer, Zanetti, Lehan, Huang and Gay}]{Robbins2014}
\bibinfo{author}{Robbins, S.J.}, \bibinfo{author}{Antonenko, I.},
  \bibinfo{author}{Kirchoff, M.R.}, \bibinfo{author}{Chapman, C.R.},
  \bibinfo{author}{Fassett, C.I.}, \bibinfo{author}{Herrick, R.R.},
  \bibinfo{author}{Singer, K.}, \bibinfo{author}{Zanetti, M.},
  \bibinfo{author}{Lehan, C.}, \bibinfo{author}{Huang, D.},
  \bibinfo{author}{Gay, P.L.}, \bibinfo{year}{2014}.
\newblock \bibinfo{title}{{The variability of crater identification among
  expert and community crater analysts}}.
\newblock \bibinfo{journal}{Icarus} \bibinfo{volume}{234},
  \bibinfo{pages}{109--131}.
\newblock \URLprefix \url{http://dx.doi.org/10.1016/j.icarus.2014.02.022},
  \DOIprefix\doi{10.1016/j.icarus.2014.02.022}.
\bibitem[{Robbins and Hynek(2012a)}]{Robbins2012a}
\bibinfo{author}{Robbins, S.J.}, \bibinfo{author}{Hynek, B.M.},
  \bibinfo{year}{2012}a.
\newblock \bibinfo{title}{{A new global database of Mars impact craters larger
  than 1 km: 2. Global crater properties and regional variations of the
  simple-to-complex transition diameter}}.
\newblock \bibinfo{journal}{Journal of Geophysical Research E: Planets}
  \bibinfo{volume}{117}, \bibinfo{pages}{1--21}.
\newblock \DOIprefix\doi{10.1029/2011JE003967}.
\bibitem[{Robbins and Hynek(2012b)}]{Robbins2012}
\bibinfo{author}{Robbins, S.J.}, \bibinfo{author}{Hynek, B.M.},
  \bibinfo{year}{2012}b.
\newblock \bibinfo{title}{{A new global database of Mars impact craters larger
  than 1km: 1. Database creation, properties, and parameters.}}
\newblock \bibinfo{journal}{Journal of Geophysical Research E: Planets}
  \bibinfo{volume}{117}, \bibinfo{pages}{1--18}.
\newblock \DOIprefix\doi{10.1029/2011JE003966}.
\bibitem[{Robbins et~al.(2018a)Robbins, Riggs, Weaver, Bierhaus, Chapman,
  Kirchoff, Singer and Gaddis}]{Robbins2018}
\bibinfo{author}{Robbins, S.J.}, \bibinfo{author}{Riggs, J.D.},
  \bibinfo{author}{Weaver, B.P.}, \bibinfo{author}{Bierhaus, E.B.},
  \bibinfo{author}{Chapman, C.R.}, \bibinfo{author}{Kirchoff, M.R.},
  \bibinfo{author}{Singer, K.N.}, \bibinfo{author}{Gaddis, L.R.},
  \bibinfo{year}{2018}a.
\newblock \bibinfo{title}{{Revised recommended methods for analyzing crater
  size-frequency distributions}}.
\newblock \bibinfo{journal}{Meteoritics and Planetary Science}
  \bibinfo{volume}{53}, \bibinfo{pages}{891--931}.
\newblock \DOIprefix\doi{10.1111/maps.12990}.
\bibitem[{Robbins et~al.(2018b)Robbins, Watters, Chappelow, Bray, Daubar,
  Craddock, Beyer, Landis, Ostrach, Tornabene, Riggs and Weaver}]{Robbins2018a}
\bibinfo{author}{Robbins, S.J.}, \bibinfo{author}{Watters, W.A.},
  \bibinfo{author}{Chappelow, J.E.}, \bibinfo{author}{Bray, V.J.},
  \bibinfo{author}{Daubar, I.J.}, \bibinfo{author}{Craddock, R.A.},
  \bibinfo{author}{Beyer, R.A.}, \bibinfo{author}{Landis, M.},
  \bibinfo{author}{Ostrach, L.R.}, \bibinfo{author}{Tornabene, L.},
  \bibinfo{author}{Riggs, J.D.}, \bibinfo{author}{Weaver, B.P.},
  \bibinfo{year}{2018}b.
\newblock \bibinfo{title}{{Measuring impact crater depth throughout the solar
  system}}.
\newblock \bibinfo{journal}{Meteoritics and Planetary Science}
  \bibinfo{volume}{53}, \bibinfo{pages}{583--637}.
\newblock \DOIprefix\doi{10.1111/maps.12956}.
\bibitem[{Ronneberger et~al.(2015)Ronneberger, Fischer and
  Brox}]{Ronneberger2015}
\bibinfo{author}{Ronneberger, O.}, \bibinfo{author}{Fischer, P.},
  \bibinfo{author}{Brox, T.}, \bibinfo{year}{2015}.
\newblock \bibinfo{title}{{U-Net: Convolutional Networks for Biomedical Image
  Segmentation}}, in: \bibinfo{editor}{Navab, N.}, \bibinfo{editor}{Hornegger,
  J.}, \bibinfo{editor}{Wells, W.M.}, \bibinfo{editor}{Frangi, A.F.} (Eds.),
  \bibinfo{booktitle}{Medical Image Computing and Computer-Assisted
  Intervention -- MICCAI 2015}, \bibinfo{publisher}{Springer International
  Publishing}, \bibinfo{address}{Cham}. pp. \bibinfo{pages}{234--241}.
\bibitem[{Salamuni{\'{c}}car et~al.(2011)Salamuni{\'{c}}car, Lonari{\'{c}},
  Pina, Bandeira and Saraiva}]{Salamuniccar2011}
\bibinfo{author}{Salamuni{\'{c}}car, G.}, \bibinfo{author}{Lonari{\'{c}}, S.},
  \bibinfo{author}{Pina, P.}, \bibinfo{author}{Bandeira, L.},
  \bibinfo{author}{Saraiva, J.}, \bibinfo{year}{2011}.
\newblock \bibinfo{title}{{MA130301GT catalogue of Martian impact craters and
  advanced evaluation of crater detection algorithms using diverse topography
  and image datasets}}.
\newblock \bibinfo{journal}{Planetary and Space Science} \bibinfo{volume}{59},
  \bibinfo{pages}{111--131}.
\newblock \DOIprefix\doi{10.1016/j.pss.2010.11.003}.
\bibitem[{Salamuni{\'{c}}car et~al.(2012)Salamuni{\'{c}}car,
  Lon{\v{c}}ari{\'{c}} and Mazarico}]{Salamuniccar2012}
\bibinfo{author}{Salamuni{\'{c}}car, G.},
  \bibinfo{author}{Lon{\v{c}}ari{\'{c}}, S.}, \bibinfo{author}{Mazarico, E.},
  \bibinfo{year}{2012}.
\newblock \bibinfo{title}{{LU60645GT and MA132843GT catalogues of Lunar and
  Martian impact craters developed using a Crater Shape-based interpolation
  crater detection algorithm for topography data}}.
\newblock \bibinfo{journal}{Planetary and Space Science} \bibinfo{volume}{60},
  \bibinfo{pages}{236--247}.
\newblock \DOIprefix\doi{10.1016/j.pss.2011.09.003}.
\bibitem[{Silburt et~al.(2019)Silburt, Ali-Dib, Zhu, Jackson, Valencia, Kissin,
  Tamayo and Menou}]{Silburt2019}
\bibinfo{author}{Silburt, A.}, \bibinfo{author}{Ali-Dib, M.},
  \bibinfo{author}{Zhu, C.}, \bibinfo{author}{Jackson, A.},
  \bibinfo{author}{Valencia, D.}, \bibinfo{author}{Kissin, Y.},
  \bibinfo{author}{Tamayo, D.}, \bibinfo{author}{Menou, K.},
  \bibinfo{year}{2019}.
\newblock \bibinfo{title}{{Lunar crater identification via deep learning}}.
\newblock \bibinfo{journal}{Icarus} \bibinfo{volume}{317},
  \bibinfo{pages}{27--38}.
\newblock \DOIprefix\doi{10.1016/j.icarus.2018.06.022},
  \href{http://arxiv.org/abs/1803.02192}{\tt arXiv:1803.02192}.
\bibitem[{Soderblom et~al.(1974)Soderblom, Condit, West, Herman and
  Kreidler}]{Soderblom1974}
\bibinfo{author}{Soderblom, L.A.}, \bibinfo{author}{Condit, C.D.},
  \bibinfo{author}{West, R.A.}, \bibinfo{author}{Herman, B.M.},
  \bibinfo{author}{Kreidler, T.J.}, \bibinfo{year}{1974}.
\newblock \bibinfo{title}{{Martian planetwide crater distributions:
  Implications for geologic history and surface processes}}.
\newblock \bibinfo{journal}{Icarus} \bibinfo{volume}{22},
  \bibinfo{pages}{239--263}.
\newblock \DOIprefix\doi{10.1016/0019-1035(74)90175-4}.
\bibitem[{Stepinski et~al.(2009)Stepinski, Mendenhall and Bue}]{Stepinski2009}
\bibinfo{author}{Stepinski, T.F.}, \bibinfo{author}{Mendenhall, M.P.},
  \bibinfo{author}{Bue, B.D.}, \bibinfo{year}{2009}.
\newblock \bibinfo{title}{{Machine cataloging of impact craters on Mars}}.
\newblock \bibinfo{journal}{Icarus} \bibinfo{volume}{203},
  \bibinfo{pages}{77--87}.
\newblock \URLprefix \url{http://dx.doi.org/10.1016/j.icarus.2009.04.026},
  \DOIprefix\doi{10.1016/j.icarus.2009.04.026}.
\bibitem[{Stepinski and Urbach(2009)}]{Stepinski2009b}
\bibinfo{author}{Stepinski, T.F.}, \bibinfo{author}{Urbach, E.R.},
  \bibinfo{year}{2009}.
\newblock \bibinfo{title}{{The First Automatic Survey of Impact Craters on
  Mars: Global Maps of Depth/ Diameter Ratio}}, in: \bibinfo{booktitle}{Lunar
  and Planetary Science Conference}, p. \bibinfo{pages}{1117}.
\newblock \DOIprefix\doi{10.2174/138920312803582960}.
\bibitem[{van~der Walt et~al.(2014)van~der Walt, Sch{\"{o}}nberger,
  Nunez-Iglesias, Boulogne, Warner, Yager, Gouillart and Yu}]{VanderWalt2014}
\bibinfo{author}{van~der Walt, S.}, \bibinfo{author}{Sch{\"{o}}nberger, J.L.},
  \bibinfo{author}{Nunez-Iglesias, J.}, \bibinfo{author}{Boulogne, F.},
  \bibinfo{author}{Warner, J.D.}, \bibinfo{author}{Yager, N.},
  \bibinfo{author}{Gouillart, E.}, \bibinfo{author}{Yu, T.},
  \bibinfo{year}{2014}.
\newblock \bibinfo{title}{{scikit-image: image processing in Python}}.
\newblock \bibinfo{journal}{PeerJ} \bibinfo{volume}{2}, \bibinfo{pages}{e453}.
\newblock \URLprefix \url{https://peerj.com/articles/453},
  \DOIprefix\doi{10.7717/peerj.453}, \href{http://arxiv.org/abs/1407.6245}{\tt
  arXiv:1407.6245}.
\bibitem[{Wang et~al.(2020)Wang, Xie, Xiao and Cui}]{Wang2020}
\bibinfo{author}{Wang, Y.}, \bibinfo{author}{Xie, M.}, \bibinfo{author}{Xiao,
  Z.}, \bibinfo{author}{Cui, J.}, \bibinfo{year}{2020}.
\newblock \bibinfo{title}{{The minimum confidence limit for diameters in crater
  counts}}.
\newblock \bibinfo{journal}{Icarus} ,
  \bibinfo{pages}{113645}\DOIprefix\doi{10.1016/j.icarus.2020.113645}.
\bibitem[{Wronkiewicz et~al.(2018)Wronkiewicz, Kerner and
  Harrison}]{Wronkiewicz2018}
\bibinfo{author}{Wronkiewicz, M.}, \bibinfo{author}{Kerner, H.R.},
  \bibinfo{author}{Harrison, T.}, \bibinfo{year}{2018}.
\newblock \bibinfo{title}{{Autonomous Mapping of Surface Features on Mars}},
  in: \bibinfo{booktitle}{AGU Fall Meeting}, pp. \bibinfo{pages}{P41D--3758}.
\bibitem[{Zhang et~al.(2018)Zhang, Liu and Wang}]{Zhang2018a}
\bibinfo{author}{Zhang, Z.}, \bibinfo{author}{Liu, Q.}, \bibinfo{author}{Wang,
  Y.}, \bibinfo{year}{2018}.
\newblock \bibinfo{title}{{Road Extraction by Deep Residual U-Net}}.
\newblock \bibinfo{journal}{IEEE Geoscience and Remote Sensing Letters}
  \bibinfo{volume}{15}, \bibinfo{pages}{749--753}.
\newblock \DOIprefix\doi{10.1109/LGRS.2018.2802944},
  \href{http://arxiv.org/abs/1711.10684}{\tt arXiv:1711.10684}.
\bibitem[{Zuo et~al.(2016)Zuo, Zhang, Li, Wang, Yu and Geng}]{Zuo2016}
\bibinfo{author}{Zuo, W.}, \bibinfo{author}{Zhang, Z.}, \bibinfo{author}{Li,
  C.}, \bibinfo{author}{Wang, R.}, \bibinfo{author}{Yu, L.},
  \bibinfo{author}{Geng, L.}, \bibinfo{year}{2016}.
\newblock \bibinfo{title}{{Contour-based automatic crater recognition using
  digital elevation models from Chang'E missions}}.
\newblock \bibinfo{journal}{Computers and Geosciences} \bibinfo{volume}{97},
  \bibinfo{pages}{79--88}.
\newblock \URLprefix \url{http://dx.doi.org/10.1016/j.cageo.2016.07.013},
  \DOIprefix\doi{10.1016/j.cageo.2016.07.013}.

\end{thebibliography}
\end{document}